\documentstyle[aps]{revtex}

\begin{document}
\twocolumn
\wideabs{

\title{A Simple Model of Liquid-Liquid Phase Transitions}
\author{
H. K. Lee\footnote{Current Address: Dept. of Physics, University of
Georgia, Athens, GA 30602, U.S.A, Computer Science and Mathematics Department, 
Oak Ridge National Laboratory, Bldg 6012, Bethel Valley Road, Oak Ridge, 
TN 37831-6367, U.S.A.
} and R. H. Swendsen \\
\footnotesize Dept. of Physics, Carnegie Mellon University, Pittsburgh, PA 15213, U.S.A.} 

\maketitle

\begin{abstract}
In recent years, a second fluid-fluid phase transition has been reported in   
several materials at pressures far above the usual liquid-gas phase           
transition.  In this paper, we introduce a new model of this behavior based   
on the Lennard-Jones interaction with a modification to mimic the different   
kinds of short-range orientational order in complex materials.  We have       
done Monte Carlo studies of this model that clearly demonstrate     
the existence of a second first-order fluid-fluid phase transition between    
high- and low-density liquid phases.                                     
\end{abstract}

{\hspace{1.5cm} \small PACS number: 64.70.Ja}
}

The most common example of a first-order phase transition is that between a 
liquid and a gas, such as boiling water.  On the other hand, while many 
transitions between different solid phases of homogeneous materials are 
also well known, it is only relatively recently that evidence of a second fluid-fluid 
phase transition has been found. In fact, liquid-liquid phase transitions 
(LLPT) have been suggested in liquid S, Ga, Se, Te, I$_2$, Cs, and Bi~\cite{ferraz}.

Stell and Hemmer~\cite{stell,hammer} showed the existence of LLPT in a one dimensional 
model with a softened hard core potential and a long-range negative attraction. 
Their work was later studied in more detail by Franzese, Malescio, et.
al.~\cite{franzese} and Sadr-Lahijany, Scala, et. al.~\cite{sadr}.

Mitus, Patashinskii and Shumilo~\cite{mitus} proposed LLPT in molten salt 
at high pressure based on a phenomenological model.
Ferraz and March suggested a similar LLPT in carbon,
with indirect experimental evidence being found by Togaya~\cite{togaya}.
Glosli and Ree~\cite{glosli} published results of a first-order
liquid-liquid phase transition in molten Carbon between two 
thermodynamically stable liquid phases.
Extensive computer simulations on models of water have supported the existence
of a LLPT in the metastable 
region~\cite{meyer,stanley,harrington,harrington1,poole,poole1,tanaka,shiratan}. 
Experimental results supporting the evidence of liquid-liquid phase
transitions in water have also been found
~\cite{mishima,soper,funel}. 
Katayama and Mizutani et. al.~\cite{katayama} found a
liquid-liquid phase transition in molten phosphorus using x-ray
diffraction. 

Our objective is to investigate the general phenomenon instead
of studying LLPT for any particular substance.
We have developed a simple model 
that exhibits a transition between high- and low-density liquids at high 
pressure. 

The behavior of our model is constructed 
to be similar to behavior seen in 
simulations of real substances such as water and carbon but without
introducing the complexity of having to simulate 
molecular orientations as in water and carbon.

To mimic the effects of local ordering, 
we have represented the different relative local orientations of 
the molecules with a spin-one-half variable. The 
interactions between particles with the same spin are given by the original 
Lennard-Jones expression, while the interactions between particles 
with opposite spin are purely repulsive.
\begin{eqnarray}
\phi_{\uparrow \uparrow} (r) &= \phi_{\downarrow \downarrow}  (r)&
= 4 \epsilon( \frac{\sigma_{l}^{12}}{r^{12}}
    - \frac{\sigma_{l}^6}{r^{6}} ) \\
\phi_{\downarrow \uparrow} (r) &= \phi_{\uparrow \downarrow } (r) &
= 4 \epsilon( \frac{\sigma_{u}^{12}}{r^{12}} )
\end{eqnarray}
Note that there are different values of $\sigma$ for like and unlike spins. 
This is an important feature of the model, and some properties, including 
the symmetries of the solid phases, are sensitive to the relative values of 
$\sigma_l$ and $\sigma_u$. The LLPT occurs when $\sigma_u$ is smaller than 
$\sigma_l$, so that by re-orienting the spins, the particles are capable of 
forming different co-ordination numbers and local structures.
We have performed most of our 
investigations for the case in which the ratio is $1/2$.

Our model can be modified to couple to a fictitious external magnetic field.
If we define the magnetic moment as the sum of all spins, then the 
Hamiltonian is given by,
\begin{equation}
H = \frac{1}{2} \sum_{i \neq j} \phi_{ij} (r_{ij}) - h \sum_{i} \sigma_i
\end{equation}
$h$ is the fictitious magnetic field which is set to zero in our simulations
and $\sigma_i = \pm 1$ is the direction of the spins of individual particles.

We used the units $T^*=k_B T/\epsilon$, $P^*=\sigma^2_l k_B P /\epsilon$. 
We have set the potential cutoffs at 3$\sigma_l$.

We performed Monte Carlo simulations of the two-dimensional 
version of our model in various ensembles.
Although our simulations are done in two dimensions, our model is not limited
to two dimensions. The simulations turned out to 
be rather difficult, and it was necessary to extend the usual techniques to 
improve efficiency. However, we did find 
clear evidence of an LLPT at high pressures. We are able to map out both
PT and P$\rho$ phase diagrams.

For our simulations of fluids, we confined the particles to a square with 
periodic boundary conditions.  For those simulations that were extended to 
include solid phases, we used parallelograms to allow us to vary the angle 
of the boundary conditions, as well as the volume of the container. The 
optimization of the Monte Carlo simulation step sizes for individual 
particle motion, changes of the volume of the box, and changes of the angle 
of the parallelogram were dynamically optimized using the Acceptance Ratio 
Method~\cite{arm}. Metropolis flips to maintain equilibrium for the spins 
associated with each particle were also carried out.

Volume changes on the dense fluid turned out to be rather inefficient 
because $\phi_{\uparrow \downarrow}$ increases very rapidly at short 
distances.
We solved this problem by introducing a cluster Monte Carlo move. The 
clusters are formed by creating bonds between particles with probability,
\begin{equation}
P(r) =
\cases{
 1 & $ r < r_{min} $ \cr
 (r_{max}-r)/(r_{max}-r_{min}) & $ r \in [r_{min},r_{max}] $\cr
 0  &otheriwse
}
\end{equation}
and then changing the size of the box by rescaling the locations of the 
centers of mass of the clusters. This cluster move is extremely effective, 
with improvements in the acceptance ratio by factors of up to $10^{10}$.

The probability of accepting a proposed cluster move is given as follows. 
Let ${\cal E}$ be the set of all edges which joins any two particles which 
are less than 3$\sigma_{l}$ apart, and let $\cal B$ be the subset of 
$\cal E $ consisting of bonds formed with the probability P(r), where $r$ 
is the edge length joining the two particles.
The probability of forming $\cal B$ is,
\begin{equation}
P({\cal B}) = \prod_{i \in {\cal B}} P(r_i) 
\prod_{j \in {\cal E} - {\cal B}} 1-P(r_j)
\end{equation}
Bond configurations need to remain invariant in order to satisfy detailed
balance. Therefore ${\cal B} = {\cal B}'$, $r_i=r_i'$ for all $r_i \in {\cal B}$
and $ r_i' \in {\cal B}'$, but $r_j \neq r_j'$ for $r_j \in {\cal E}-{\cal B}$
and $r_j' \in {\cal E'}-{\cal B'}$. Therefore,
\begin{equation}
{\prod_{i \in {\cal B}} P(r_i)}/{\prod_{i' \in {\cal B'}} P(r_i')} = 1 
\end{equation}
and,
\begin{equation}
  \frac{P({\cal B'})}{P({\cal B})}  
 =   \prod_{j' \in {\cal E'-B'}} (1-P(r_{j'}))
 /\prod_{j \in {\cal E-B}} (1-P(r_{j}))
\end{equation}
Hence the acceptance probability is,
\begin{eqnarray}
\lefteqn {A(\varpi \rightarrow \varpi')  =}  \\
& &\mbox{min}\left(1,
    e^{-\beta ( \Delta E + p \Delta V)} \frac{V'^{N}}{V^N}
    \frac{\prod_{i \in \cal{E'}-\cal{B'}} [1-P(r_i)]}{
    \prod_{i \in \cal{E}-\cal{B}} [1-P(r_i)]}
    \right)  \nonumber
\end{eqnarray}

Another new move that has proven extremely effective is to form clusters of 
nearby particles that are less than $1/\sqrt{\rho sin(\pi/3)}$ apart 
and attempt to flip the spins of all particles in a 
cluster.

For canonical ensemble simulations using two boxes in equilibrium with each 
other, the total volume is conserved~\cite{panagiotopoulos}. An 
additional Monte Carlo move is 
introduced to transfer particles between the boxes.

Thermodynamic quantities were calculated from simulations in the constant 
pressure ensemble over relatively large pressure and temperature domains 
using the Multiple Histogram Method~\cite{ferrenberg}. The coexistence 
curve is mapped out by tracing the ridge line of the isothermal 
compressibility.
The position of the tricritical point was estimated using finite size scaling.
The P$\rho$ diagram was determined by combining other data with the results 
of simulations using the canonical ensemble~\cite{panagiotopoulos}. 

\begin{figure} 
\begin{picture}(0,370)(0,0)
\put(0,380){\includegraphics{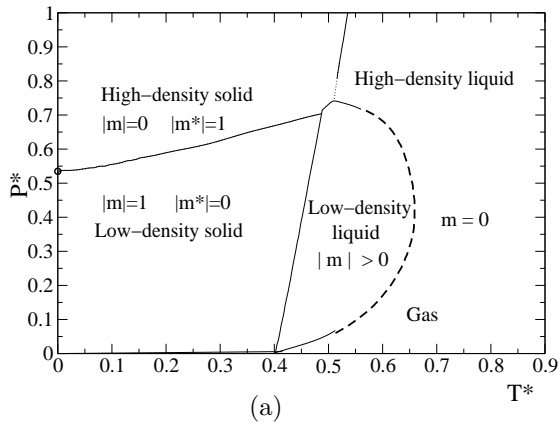}}
\put(0,197){\includegraphics{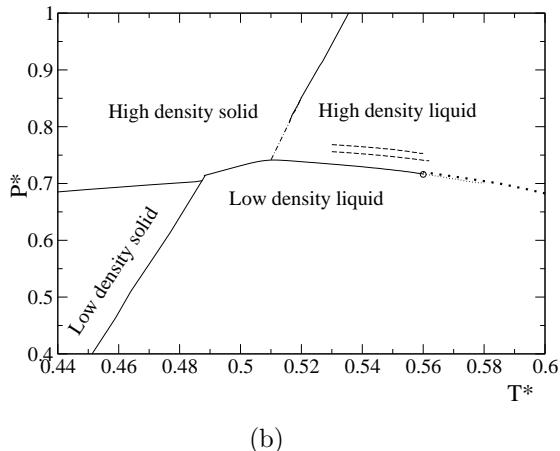}}
\put(100,5){(b)}
\put(100,200){(a)}
\end{picture}
\caption{\small 
  (a) PT diagram.
Solid lines show coexistence curves
obtained from our simulations with 160 particles.
$\circ$ marks the calculated solid-solid transition point at T=0.
Dotted lines show our estimate for the coexistence curves. The thick
dashed line represents the locus of Curie points for the second-order
magnetic transitions. $|\mbox{m}|$ and $|\mbox{m}^*|$ are the magnetic
and anti-magnetic order parameters respectively. 
 (b) PT diagram in the region of interest, the circle 
marks
the tricritical point and 
dotted line shows the peak of the isothermal
compressibility beyond tricritical point. Solid lines
show coexistence curves obtained from our simulations with 160 particles.
The dashed line shows the extent of
curve shifting due to the finite size effect. The upper dashed line corresponds
to a system with 480 particles and the lower dashed line corresponds to
a system with 320 particles.
}
\label{pt}
\end{figure}

The phase diagram for our model is shown in figure (\ref{pt}).  At low 
pressures its behavior is virtually identical to that of the usual 
Lennard-Jones model.  At higher pressures, the coexistence curve for the 
LLPT is shown projecting into the fluid region of the phase diagram.  This
line of first-order phase transitions actually ends in a tricritical point,
which separates it from a line of critical points associated with the ordering
of the ``spins'' used to define the model.  This line joins the liquid-gas
and liquid-liquid first-order coexistence curves at the tricritical points.
However, it is only our access to the ``magnetic'' degrees of 
freedom in the model that make it obvious that the LLPT curves end in 
tricritical rather than critical points.  If we did not have access to
the magnetic degrees of freedom, rather careful measurement of the 
exponents characterizing the divergences at the end of the line of 
first-order transitions would be necessary to distinguish the two cases.
It is not completely clear whether a fictitious ordering field in 
a more elaborate model would not have the same consequence of producing
tricritical points.

The apparent discontinuity of the coexistence curves between the 
high-density solid, the low-density solid and the low-density liquid
is an artifact of the accuracy of our simulations. These lines are
obtained from the Multiple Histogram extrapolations in opposite 
directions. The fact that these two lines do not meet exactly,
but terminated very close to each other shows the accuracy of
our simulations.

To determine the liquid-liquid coexistence curve in the PT diagram, 
simulations were done with 160 particles at constant pressure, 
using larger systems to check the results. Sets of histograms were 
collected at $T^*$=0.53, 0.55, and 0.565 over a range of pressures covering 
across the coexistence region.
The system was equilibrated for 30,000 MCS/P before beginning 
to take data for another 300,000 MCS/P.

To determine the tricritical point for the liquid-liquid phase transition, 
additional simulations for 160, 240, 320, and 480 particles were done at 
constant pressures at $T^*$=0.55 and 0.565.
The Monte Carlo simulations used a total    
of 300,000 MCS/P  for the 160 particle systems and typical equilibration     
runs of 30,000 MCS/P.  For larger systems, longer runs were used.  For the    
480 particle system, a total of 1,200,000 MCS/P were used with typical        
equilibration runs of 100,000 MCS/P.                                          

To determine the liquid-solid coexistence curve, simulations were performed 
at $P^*$=0.5, 0.6, 0.78 and 0.9 over a range of temperatures covering the 
solid-liquid coexistence region. The system was equilibrated for 60,000 
steps per particle before collecting data for another 600,000 steps per 
particle. 
 
\begin{figure}[h]
\begin{picture}(10,400)(0,0)
\put(-150,-60){\includegraphics{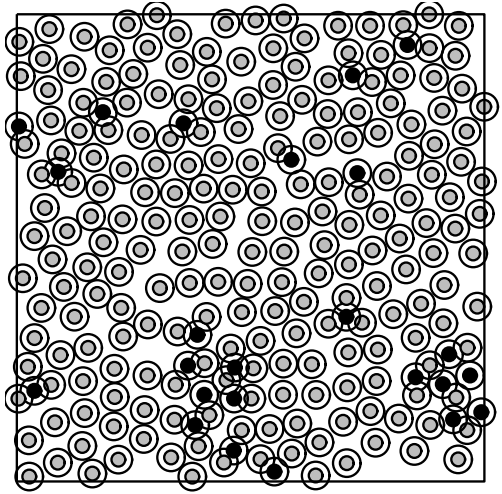}}
\put(-100,-190){\includegraphics{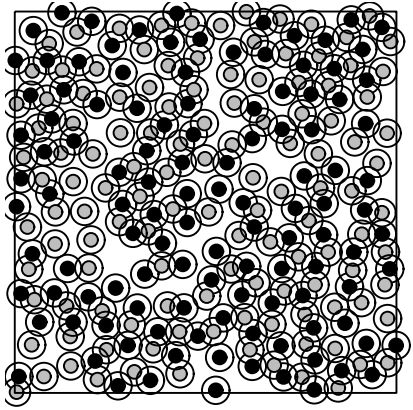}}
\put(100,10){(b)}
\put(100,200){(a)}
\end{picture}
\caption{\small (a,b) Snapshots of a system of 240 particles in the low-density(a) and
high-density(b) liquid state coexisting at $T^*$=0.55,
$P^*$=0.75, $\rho^*$=0.833 and 1.25 respectively.}
\label{configpic}
\end{figure}

The phase transition between high- and low-density phases is strongly 
first-order. Figure (\ref{configpic}a,b) shows snap shots of high- and 
low-density liquids near the coexistence curve for a system with 240 
particles.
The low-density liquid is dominated by parallel nearest-neighbor 
interactions with a stronger core repulsion and with co-ordination 
number six. The high-density liquid is dominated by an anti-parallel 
nearest-neighbor interaction that 
allows the particles to come closer together and with co-ordination 
number three.  The differences in local 
orderings for low- and high-density liquids stabilize the liquids and 
create the possibility of liquid-liquid phase transitions.

\begin{figure}[h]
\begin{picture}(10,130)(0,0)
\put(-20,185){\includegraphics{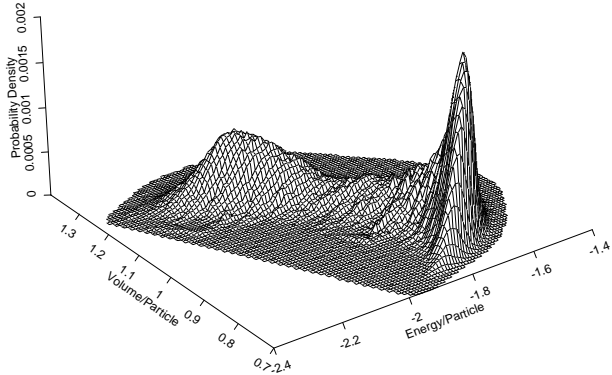}}
\end{picture}
\caption{\small Probability density function for 240 particles coexisting 
at
  $T^*$=0.54 and $P^*$=0.75.}
\label{pev}
\end{figure}

The volume-energy probability density function
(figure (\ref{pev})) has a saddle point that is typical of a first-order 
transition.
The coexistence curve for the LLPT has a negative slope, as shown in the PT 
diagram in figure (\ref{pt}), reflecting the higher entropy of the 
high-density liquid.
The sharp bend of the liquid-solid coexistence curve at the triple point 
found from the multiple-histogram analysis is consistent with our 
observation of the liquid-liquid phase transition. 

\begin{figure}[h]
\begin{picture}(0,180)(0,0)
\put(0,190){\includegraphics{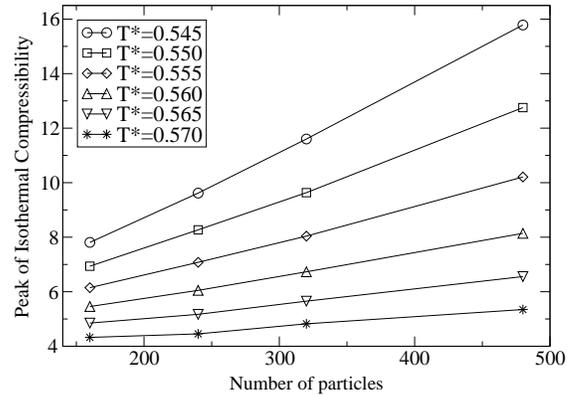}}
\end{picture}
\caption{\small Plot of peak values of isothermal compressibility
vs system size. The tricritical point is near $T^*$=0.56.}
\label{fsi}
\end{figure}

To find the tricritical point, we use the fact that peak values of the 
isothermal compressibility grow as ${\cal O}(N)$ at the coexistence curve 
and as ${\cal O}(N^{0.925})$~\cite{den,neihuis,pearson,neihuis1,den1,neihuis2}
at the tricritical point, assuming that the 
transition is in the expected two-dimensional Ising class.
Simulations were done with N=160, 240, 320 and 480 near the tricritical point.
Peak values of the isothermal compressibility are then plotted against 
system sizes for various temperatures.
Figure (\ref{fsi}) shows the size dependence of the isothermal 
compressibility. Although it is hard to determine the exact location of the 
tricritical point, the data suggests that it is located at $T^*=0.56 \pm 
0.01$.

\begin{figure}[!h]
\begin{picture}(0,220)(0,0)
\put(0,200){\includegraphics{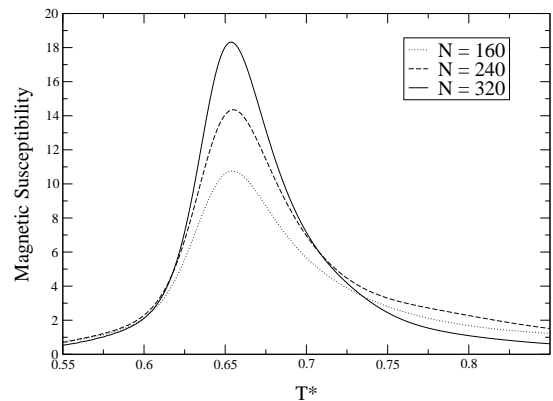}}
\end{picture}
\caption{\small Magnetic susceptibility plotted against $T^*$ at
$P^*=0.5$ and $h=0$.}
\label{chi}
\end{figure}

Figure \ref{chi} shows the growing divergence of the magnetic susceptibility 
for 160, 240 and 320 particles.
These magnetic susceptibility are plotted at $P^*=0.5$ which
is far away from the liquid-liquid coexistence curve and the
gas-liquid coexistence curve.

\begin{figure}[h]
\begin{picture}(0,180)(0,0)
\put(0,190){\includegraphics{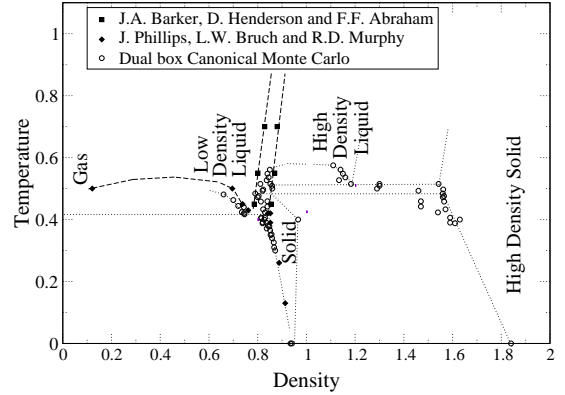}}
\end{picture}
\caption{\small $\rho$-T diagram. Points obtained by simulation using the 
canonical ensemble with 160 particles in two boxes. 
Dashed lines show the coexistence region 
of the pure Lennard-Jones system reported by Barker, Henderson, Abraham
and Phillips, Bruch, Murphy. Dotted lines show the coexistence region of 
our model. All lines are drawn to guide the eye.}
\label{rhot}
\end{figure}

The $\rho$-T diagram (figure (\ref{rhot})) was mapped out using simulations 
in the canonical ensemble using dual boxes to simulate coexistence without 
the inconvenience of an interface between the 
phases~\cite{panagiotopoulos}.
The low-density 
liquid-gas coexistence region is only slightly lower than that of pure 
Lennard-Jones~\cite{barker,phillips} due to the small effect of the differences in 
the models at low densities.

\begin{figure}[h]
\begin{picture}(10,350)(0,0)
\put(-60,80){\includegraphics{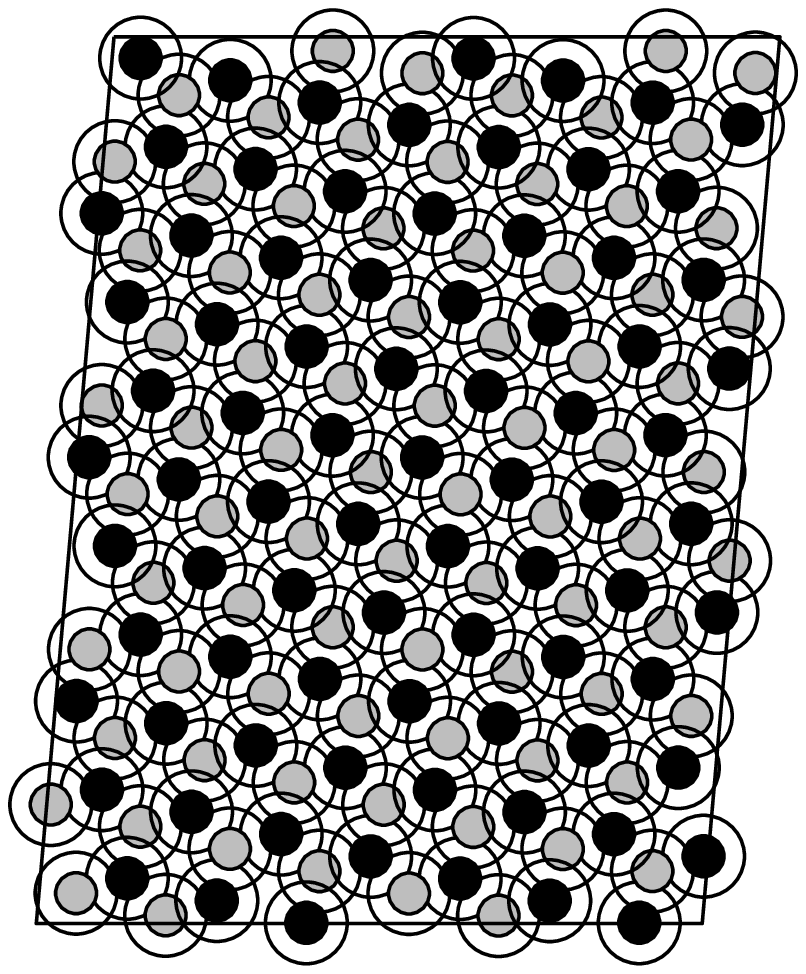}}
\put(-80,-180){\includegraphics{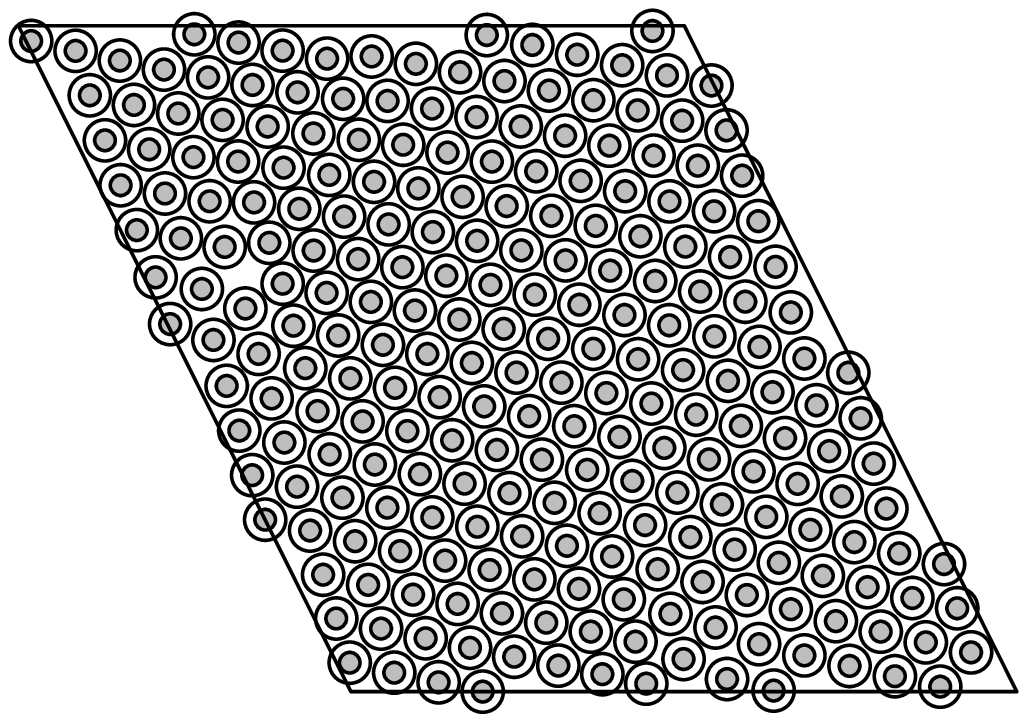}}
\put(80,210){(a)}
\put(80,10){(b)}
\end{picture}
\caption{\small (a,b) The high-density(a) and
low-density(b) solid states, notice a vacancy in the low-density
solid state.}
\label{solid}
\end{figure}

Figure(\ref{solid}a) shows a snapshot of the high-density crystalline state, 
which has three-fold symmetry and zero magnetic moment.  The low-density 
crystalline state (figure(\ref{solid}b)) is hexagonal close packed with 
uniform spin.
At zero temperature, we can calculate the location of the transition 
between the two solid phases to arbitrary precision. Its location is 
determined to be $P^*=0.534819 \pm 10^{-7}$, with $\rho^* = 1.82646$ and 
$\rho^*= 0.943122$  for the high- and low-density solids respectively. For 
comparison, $\rho^* = 0.934721$ at $T^*=0$ and $P^*=0$, so the low-density 
solid phase changes its density very little up to the boundary of the 
high-density phase.

\begin{figure}[h]
\begin{picture}(10,150)(0,0)
\put(-100,-220){\includegraphics{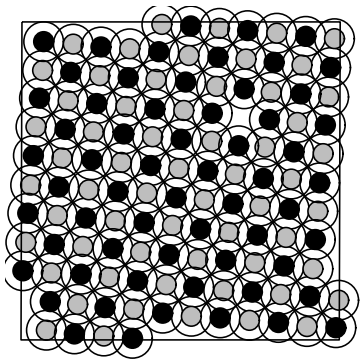}}
\end{picture}
\caption{\small Four fold solids form using a model with $\tau = 0.6$.}
\label{4solid}
\end{figure}

It was found that symmetry of the solid phase is highly dependent on the 
ratio $\tau = \sigma_{u}/\sigma_{l}$.
Figure (\ref{4solid}) shows a solid with four-fold symmetry obtained from a
simulation with $\tau = 0.6$.
We believe that the full range of solid phases is 
quite rich for this model.

We have developed a relatively simple model that demonstrates a 
liquid-liquid phase transition between high- and low-density phases. 
Although our simulations are two dimensional, our model is not restricted to
two dimensions and we see no reasion to believe that a three dimensional
simulation with our model would produce significantly different behavior. By 
comparing the behavior of our model with the properties of real systems, we 
hope to learn which properties are generic and which depend on details of 
a particular material. One immediate point 
of interest is the negative slope of the liquid-liquid coexistence line in 
our model, which reflects the high entropy in the high-density phase. This 
feature is, indeed, found in most materials that exhibit an LLPT. However, 
this does not appear to be universal, since the coexistence curve of molten 
carbon reported by Glosli and Ree~\cite{glosli} has a positive slope.  At 
the present time, this difference in materials properties is not understood.

We would like to thank Prof. Robert Griffiths for his suggestions and 
helpful comments. We also would
like to acknowledge the support from the Pittsburgh Supercomputing Center.

\end{document}